# The Random Forest Model for Analyzing and Forecasting the US Stock Market in the Context of Smart Finance

Jiajian Zheng[1,*]

Bachelor of Engineering
Guangdong University of Technology
ShenZhen,CN
`im.jiajianzheng@gmail.com`

Duan Xin[1]

Accounting
Sun Yat-Sen University
HongKong
`duanxin12314057@gmail.com`

Qishuo Cheng[2]

Department of Economics
University of Chicago
Chicago, IL, USA
`qishuoc@uchicago.edu`

Miao Tian[3]

Master of Science in Computer Science
San Fransisco Bay University
Fremont CA, USA
`miao.hnlk@gmail.com`

Le Yang[4]

Master Science in Computer Information Science
Sam Houston State University
Huntsville，TX，USA
`wesleyyang96@gmail.com`

**Abstract.** The stock market is a crucial component of the financial market, playing a vital role in wealth accumulation for investors, financing costs for listed companies, and the stable development of the national macroeconomy. Significant fluctuations in the stock market can damage the interests of stock investors and cause an imbalance in the industrial structure, which can interfere with the macro level development of the national economy. The prediction of stock price trends is a popular research topic in academia. Predicting the three trends of stock pricesrising, sideways, and falling can assist investors in making informed decisions about buying, holding, or selling stocks. Establishing an effective forecasting model for predicting these trends is of substantial practical importance. This paper evaluates the predictive performance of random forest models combined with artificial intelligence on a test set of four stocks using optimal parameters. The evaluation considers both predictive accuracy and time efficiency.

---

[1] *These authors contributed equally to this work and should be considered co-first authors.*

# 1   Introduction

Stocks and stock markets have existed for centuries. The prediction of stock price movements has been a longstanding goal for investors. In the past, when analysing listed companies, the focus was mainly on production and operational conditions, financial conditions, and technical indicators of stock trading. Even the psychology and behaviour of investors were studied. However, all analytical approaches were highly dependent on the subjective experience of the analyst, which goes against the principle of objectivity. Predicting stock price trends has always been a challenging task due to the multitude of listed companies across various industries and the complex nature of the stock market.

However, with the continuous progress of computer technology, breakthroughs in Artificial Intelligence (AI) have been achieved, particularly in intelligent finance, and are rapidly developing. Machine learning is a practical direction of AI technology that focuses on creating algorithms capable of learning from data and improving their accuracy. The algorithms are trained to discover patterns and laws in large amounts of data, enabling them to make decisions.

and predictions about new data. As data processing capabilities increase, the accuracy of the algorithm's decisions and predictions also improves. The emergence of AI technology has introduced new possibilities for analysing and predicting stock price trends. By utilising machine learning techniques, such as random forests, for stock price analysis, algorithms can be developed to assist or even guide trading decisions. This area of research shows great promise.

# 2   Related work

Machine learning in the fields of finance, stocks, etc., is a computer science approach that helps investors and financial institutions make more informed decisions by using algorithms and statistical models to analyze historical and realtime data, identify pat-terns and trends from it in order to predict the movements of financial markets, fluctuations in stock prices, and formulate investment strategies.

## 2.1 Machine learning and stock forecasting

The use of computer technology has expanded the range of investment strategies available to investors, with machine learning now widely employed in the capital market. Machine learning involves using machines to imitate human thinking processes and habits, enabling predictions and decisions to be made based on existing data. Currently, machine learning is widely used in various fields such as face recognition, intelligent investment advisory, and natural language processing. Machine learning can be classified into two categories: unsupervised learning and supervised learning. Supervised learning involves learning from existing markers, where accurate classified information is already available. In the analysis and prediction of stock data using machine learning, various algorithms are commonly used, which are based on machine learning methods. In the analysis and prediction of stock data using machine learning, various algorithms are commonly used, which are based on machine learning methods. In the analysis and prediction of stock data using machine learning, various algorithms are commonly used, which are based on machine learning methods. These methods encompass a range of techniques. The following machine learning algorithms were used: multivariate linear regression, multivariate polynomial regression, decision tree, random forest, support vector machines (SVM-NuSVR), and deep learning based on neural networks.

## 2.2 Random forest

The random forest algorithm is a nonlinear model that integrates multiple decision trees into a forest. To understand random forests, it is important to consider two key points: random sampling and majority voting. Specifically, for each decision tree, the training set is randomly selected from the entire sample set. This paper employs decision tree classification using the technical index in the feature matrix X as the standard.The decision trees make independent predictions, and the random forest combines these predictions through voting, with the most votes determining the final prediction. This approach prevents overfitting of a single decision tree.Then, a simple majority vote is taken as the classification result of the sample. Finally, the ratio of the number of misclassifications to the total number of samples is taken as the out-of-bag misclassification rate of the random forest.
To construct a Random Forest model, the first step is to set the number of trees using the `n_estimators` parameter, for example, `n_estimators=10`. In Random Forest, each tree is built independently to ensure diversity, which is achieved through a process called bootstrap sampling.For instance, suppose we have an initial list of samples ['a', 'b', 'c', 'd', 'f']. The first sample consisted of ['a', 'a', 'c', 'f', 'b'], while the second sample consisted of ['d', 'a', 'f', 'c', 'd']. Two independent bootstrap samples were used to build separate decision trees, which contributed to the ensemble of trees forming the Random Forest.

In conclusion to predict the trend of stock prices using a random forest model, it is necessary to average the prediction probability of all trees and select the category with the highest probability as the prediction result. In this case, a random forest consisting of five trees will be applied to the two moons dataset:

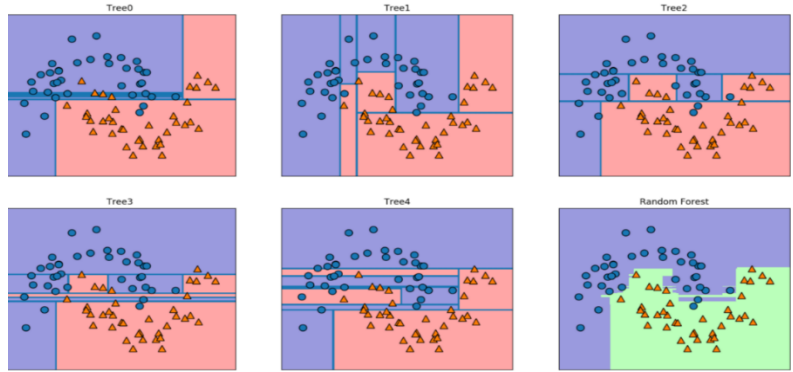

**Fig. 1.** Random forest model with five decision variables

As can be observed from the model above, the decision boundaries of these five trees differ significantly. Each tree made some errors because some of the training points shown here were not included in the training set for these trees. This is mainly due to the selfsampling random forest being smaller than the overfit of a single tree, resulting in more intuitive decision boundaries. However, in actual problemsolving, we will use a larger number of trees (hundreds of thousands) to achieve a smoother interface.
In summary, a random forest is a classifier composed of multiple decision trees built randomly. The output category is determined by the mode of the categories output by each tree. The randomness is mainly reflected in two aspects:
(1) During the training of each tree, a dataset of size N is selected from all training samples, which may contain duplicates, for training purposes. This is known as bootstrap sampling.
(2) At each node, a subset of features is randomly selected to calculate the optimal segmentation method.
Therefore, this paper uses random forest regression to predict stock returns in the stock market.

## 3 Methodology

### 3.1 Data collection and preprocessing

This model predicts the stock trends of Apple, Samsung, and GE for the next 30, 60, and 90 days based on collected data from approximately 7,000 trading days. The author preprocessed the stock price data using the exponential smoothing method to remove noise and reveal the actual patterns in the historical data:

$$S_0 = Y_0$$
$$for\ t > 0, S_t = a * Y_t + (1 - a) * S_{t-1} \quad (1)$$

Alpha, which ranges between 0 and 1 and is typically closer to 1, assigns greater weight to recent data. This is because recent movements are somewhat more consistent, so greater weight is placed on the most recent data.

### 3.2 Feature extraction

**Label setting**

$$target_i = Sign(close_{i+d} - close_i) \quad (1)$$

In formula (2), d is the predicted time window and Sign is the symbolic function. When the value of targeti is 1, it means that at the moment i, the closing price after d is higher than the closing price today, that is, the stock will rise after d; Vice versa.

target is also the target that the model needs to predict

**Classification feature**

Technical indicators are important signals used to judge bear and bull in stock analysis. This study mainly uses six technical indicators as classification criteria and lets random forest model learn these characteristics. Indicators are listed as follows:

. **Table 1.** Stock price trend forecast in six technical indicators as classification

| RSI | Stochastic Oscillator | Williams %R |
|---|---|---|
| MACD | Price Rate of Change. | On Balance Volume |

### 3.3 Prediction model

Prior to establishing the model, linear divisible tests were conducted on two types of data: rise or fall. However, it was discovered that the stock trend prediction problem was not linearly divisible. This was due to the significant overlap of the convex hull when projected into twodimensional space. Therefore, algorithms related to linear discriminant analysis, such as SVM, were deemed inapplicable. As a nonlinear algorithm, the random forest can avoid the situation mentioned above and has significant applications in predicting stock trends.

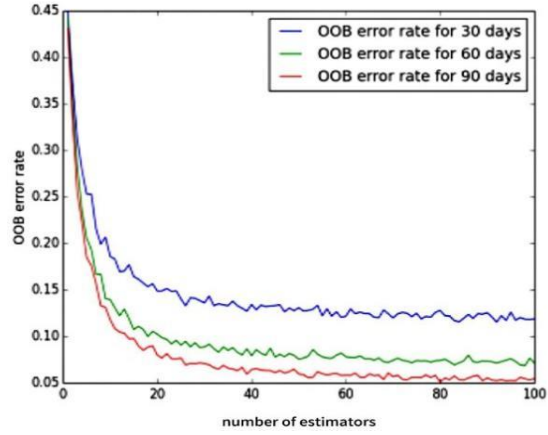

**Fig. 2.** Linear separable test results

### 3.4 Establishment of random forest hair model

The Random Forest model was used to predict stock price trends based on the linear decision described above. Apple's stock price was used to forecast trends for 30, 60, and 90 days in the future. The accuracy of the model increases and tends to converge with the increase of decision trees in the model. Furthermore, the accuracy of the model increases with longer forecast time windows.

**Table 2.** OOB error Specific result

| Trading Period (Days) | No. of Trees | Sample Size | OOB error |
|---|---|---|---|
| 30 | 5 | 6590 | 0.241729893778 |
| 30 | 25 | 6590 | 0.149165402124 |
| 30 | 45 | 6590 | 0.127617602428 |
| 30 | 65 | 6590 | 0.123672230653 |
| 60 | 5 | 6545 | 0.198472116119 |
| 60 | 25 | 6545 | 0.0890756302521 |
| 60 | 45 | 6545 | 0.0786860198625 |
| 60 | 65 | 6545 | 0.0707410236822 |
| 90 | 5 | 6500 | 0.191384615385 |
| 90 | 25 | 6500 | 0.0741538461538 |
| 90 | 45 | 6500 | 0.0647692307692 |

Table 2 shows that OOB errors decrease gradually and converge, reaching a steady state when the number of decision trees exceeds 45. The OOB error decreases slightly as the forecast time window increases.

### 3.5 ROC curve results of stock price prediction model

ROC curve (receiver operating characteristic curve) plays an important role in the evaluation of training models and stock prediction. In the field of machine learning and stock forecasting, the ROC curve is primarily used to evaluate the performance of classification models, especially in the ability to distinguish between different categories (such as rising or falling stock prices). The curve compares the performance of True Positive Rate (TPR) and False Positive Rate (FPR) at different thresholds to show the performance of the model across all possible classification thresholds. In stock forecasting, a highperformance model will be near the top left corner of the ROC curve, indicating that it can effectively distinguish between rising and falling stock prices, thus helping investors or analysts to make more accurate forecasting decisions. The area under the ROC curve (AUC) can be used as a single measure of the overall classification performance of the model, where a higher AUC value indicates that the model has better discriminating ability.

The study compares the Random Forest algorithm with SVM, logistic regression, Gaussian discriminant analysis, quadratic discriminant analysis, and other models to determine its superiority. In comparison, the accuracy rate of SVM, logistic regression, Gaussian discriminant analysis, and quadratic discriminant analysis models ranged from 30% to 80%.The random forest algorithm achieved the highest accuracy rate of 80-99%, which remained stable at about 98% when the prediction time window was greater than 60 days.

Therefore,the experimental model can predict the stock price of Apple, Samsung and General Electric with an accuracy of more than 85%, and the random forest model has excellent performance.

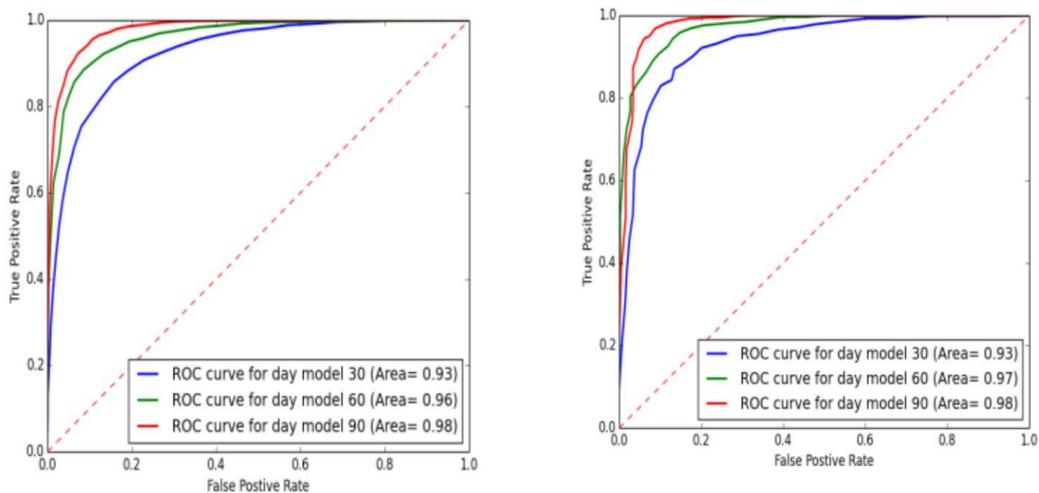

**Fig. 3.** ROC curve of Apple and Samsung stock price prediction accuracy

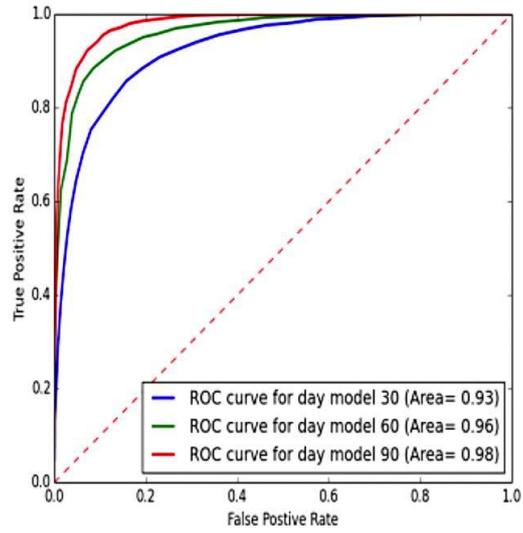

**Fig. 4.** ROC curve of GE stock price prediction accuracy

### 3.6 Comparison of model results

In order to prove the superiority of random forest algorithm, it needs to be compared with SVM, logistic regression, Gaussian discriminant analysis, quadratic discriminant analysis and other models.

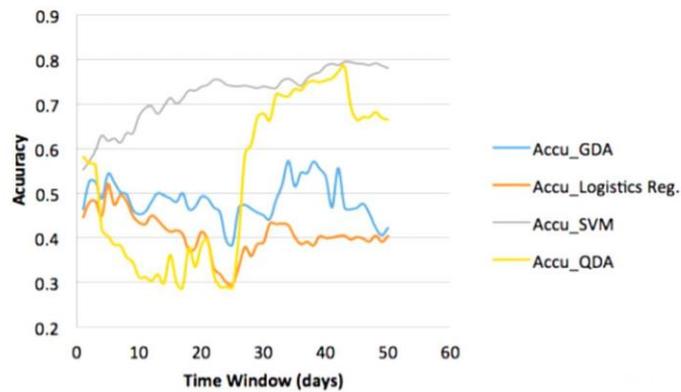

**Fig. 5.** Results of four monitoring algorithms

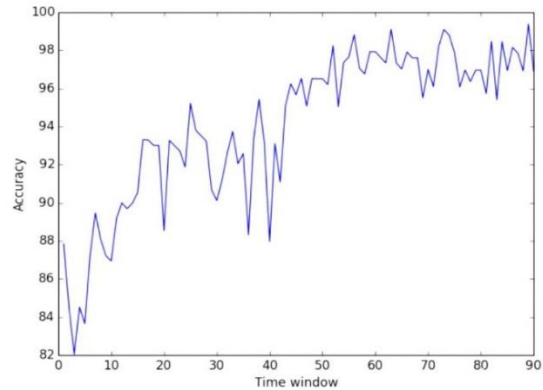

**Fig. 6.** Random forest algorithm results

This paper analyses the application of the random forest algorithm combined with artificial intelligence in predicting stock price trends in the context of smart finance. The results show that using the random forest classifier to predict longterm stock trends has achieved remarkable accuracy. The article uses data sets from Apple, Samsung and General Electric to show that the predictive accuracy of the random forest model can reach 85 to 95 per cent. Increasing the number of decision trees in a random forest can lead to more stable results. This research can inform the design of equity investment strategies.

## 4      Conclusion

This paper utilizes artificial intelligence and deep learning methods, along with the random forest model, to predict trends in the US stock market within the context of intelligent finance. The random forest model is a flexible and powerful machine learning algorithm that efficiently predicts changes in stock prices (upward, sideways, or downward) by integrating multiple decision trees. The paper collects and preprocesses stock market data, extracts key features, and uses the random forest model for classification and prediction. The experimental results demonstrate that the random forest model is highly accurate and stable in predicting stock price trends, particularly in the long term.
Additionally, the results indicate that the random forest model outperforms these methods in predicting stock price trends. These findings not only demonstrate the practical potential of the random forest in stock price prediction but also provide new insights for future financial market analysis and decisionmaking. This will provide investors and market analysts with more accurate and efficient tools, improving investment strategies and decisionmaking processes.

## 5   Acknowledgment

The research in this paper is informed by Liu Bo and his collaborators' paper 'Integration and Performance Analysis of Artificial Intelligence and Computer Vision Based on Deep Learning Algorithms' (arXiv preprint arXiv:2312.12872). Their research on the application of AI and deep learning algorithms to computer vision provides a valuable reference for our analysis and methodology. We acknowledge Liu Bo and his team for their contributions. Their advanced research results have significantly influenced the writing and research direction of this paper.
link: Liu, Bo, et al. "Integration and Performance Analysis of Artificial Intelligence and Computer Vision Based on Deep Learning Algorithms." arXiv preprint arXiv:2312.12872 (2023).

## 6   Reference

1. Markowitz, H. M. Portfolio Selection [J]. The Journal of Finance, 1952, 7(1):77-91.
2. Tian, M., Shen, Z., Wu, X., Wei, K., & Liu, Y. (2023). The Application of Artificial Intelligence in Medical Diagnostics: A New Frontier. Academic Journal of Science and Technology, 8(2), 57-61.7
3. Liu, Bo & Zhao, Xinyu & Hu, Hao & Lin, Qunwei & Huang, Jiaxin. (2023). Detection of Esophageal Cancer Lesions Based on CBAM Faster R-CNN. Journal of Theory and Practice of Engineering Science. 3. 36-42. 10.53469/jtpes.2023.03(12).06.
4. Shen, Z., Wei, K., Zang, H., Li, L., & Wang, G. (2023). The Application of Artificial Intelligence to The Bayesian Model Algorithm for Combining Genome Data. Academic Journal of Science and Technology, 8(3), 132-135.2
5. Liu, B., Zhao, X., Hu, H., Lin, Q., & Huang, J. (2023). Detection of Esophageal Cancer Lesions Based on CBAM Faster R-CNN. *Journal of Theory and Practice of Engineering Science*, *3*(12), 36–42.
6. Liu, Bo, et al. "Integration and Performance Analysis of Artificial Intelligence and Computer Vision Based on Deep Learning Algorithms." *arXiv preprint arXiv:2312.12872* (2023).
7. Li, Linxiao, et al. "Zero-resource knowledge-grounded dialogue generation." Advances in Neural Information Processing Systems 33 (2020): 8475-8485.
8. Liu, Yuxiang, et al. "Grasp and Inspection of Mechanical Parts based on Visual Image Recognition Technology." Journal of Theory and Practice of Engineering Science 3.12 (2023): 22-28.
9. K. Tan and W. Li, "Imaging and Parameter Estimating for Fast Moving Targets in Airborne SAR," in *IEEE Transactions on Computational Imaging*, vol. 3, no. 1, pp. 126-140, March 2017, doi: 10.1109/TCI.2016.2634421.
10. Zong, Yanqi, et al. "Improvements and Challenges in StarCraft II Macro-Management A Study on the MSC Dataset". Journal of Theory and Practice of Engineering Science, vol. 3, no. 12, Dec. 2023, pp. 29-35, doi:10.53469/jtpes.2023.03(12).05.
11. Tian, Miao, et al. "The Application of Artificial Intelligence in Medical Diagnostics: A New Frontier." Academic Journal of Science and Technology 8.2 (2023): 57-61.
12. Wan, Weixiang, et al. "Development and Evaluation of Intelligent Medical Decision Support Systems." Academic Journal of Science and Technology 8.2 (2023): 22-25.